\documentclass[preprint,11pt]{aastex}

\newcommand{\tsph}{TreeSPH}
\newcommand{\lya}{Ly$\alpha$\ }

\newcommand\plotonesmall[1]{%
 \centering 
 \leavevmode 
 \columnwidth=.45\columnwidth 
 \includegraphics[width={\columnwidth}]{#1}%
}%

\def\ie{i.e.\ }

\def\be{\begin{equation}}
\def\ee{\end{equation}}
\def\bdm{\begin{displaymath}}
\def\edm{\end{displaymath}}
\def\ea{{et al.\/} }

\def\kmpersec{km s$^{-1}$}
\def\kms{{\rm km}\;{\rm s}^{-1}}

\def\lta{\mathrel{\rlap{\lower 3pt\hbox{$\mathchar"218$}}
    \raise 2.0pt\hbox{$\mathchar"13C$}}}
\def\hinv{$h^{-1}$}

\newbox\grsign \setbox\grsign=\hbox{$>$} \newdimen\grdimen \grdimen=\ht\grsign
\newbox\simlessbox \newbox\simgreatbox
\setbox\simgreatbox=\hbox{\raise.5ex\hbox{$>$}\llap
     {\lower.5ex\hbox{$\sim$}}}\ht1=\grdimen\dp1=0pt
\setbox\simlessbox=\hbox{\raise.5ex\hbox{$<$}\llap
     {\lower.5ex\hbox{$\sim$}}}\ht2=\grdimen\dp2=0pt
\newcommand{\simgt}{\mathrel{\copy\simgreatbox}}
\newcommand{\simlt}{\mathrel{\copy\simlessbox}}

\newcommand\cdunits{{\rm cm}^{-2}}

\begin{document}

\title{THE INFLUENCE OF $\Omega_{\rm baryon}$ ON HIGH-REDSHIFT STRUCTURE}

\author{Jeffrey P. Gardner$^{1,2,3}$, Neal Katz$^{4}$, 
Lars Hernquist$^{5}$, and David H. Weinberg$^6$}
\affil{E-mail:  gardner@phyast.pitt.edu, nsk@kaka.astro.umass.edu,
 lars@cfa.harvard.edu, dhw@astronomy.ohio-state.edu}
\footnotetext[1]
{Department of Physics and Astronomy, University of Pittsburgh,
Pittsburgh, PA 15260}
\footnotetext[2]
{Institute of Astronomy, Madingley Road, Cambridge, CB3 0HA, UK}
\footnotetext[3]
{NSF-NATO Postdoctoral Fellow}
\footnotetext[4]{University of Massachusetts, 
Department of Astronomy, Amherst, MA 01003-4525}
\footnotetext[5]
{Department of Astronomy, Harvard University, Cambridge, MA 02138}
\footnotetext[6]
{Ohio State University, Department of Astronomy, Columbus, OH 43210}

\keywords{galaxies: formation, large scale
structure of the Universe}

\begin{abstract}
We analyze high-redshift structure in three hydrodynamic simulations that have
identical initial conditions and cosmological parameters and differ only
in the value of the baryon density parameter, $\Omega_b=0.02$, 0.05, 0.125.
Increasing $\Omega_b$ does not change the fraction of baryons in the diffuse 
(unshocked) phase of the intergalactic medium, but it increases cooling rates
and therefore transfers some baryons from the shocked intergalactic phase to
the condensed phase associated with galaxies. Predictions of Lyman-alpha forest
absorption are almost unaffected by changes of $\Omega_b$ provided that the UV
background intensity is adjusted so that the mean opacity of the forest matches 
the observed value.  The required UV background intensity scales as 
$\Omega_b^{1.7}$, and the higher photoionization rate increases the gas 
temperature in low density regions. Damped Lyman-alpha absorption and Lyman 
limit absorption both increase with increasing $\Omega_b$, though the impact is 
stronger for damped absorption and is weaker at $z=4$ than at $z=2-3$. The mass
of cold gas and stars in high-redshift galaxies increases faster than $\Omega_b$
but slower than $\Omega_b^2$, and the global star formation rate scales 
approximately as $\Omega_b^{1.5}$.  In the higher $\Omega_b$ models, the 
fraction of baryonic material within the virial radius of dark matter halos is 
usually higher than the universal fraction, indicating that gas dynamics and 
cooling can lead to over-representation of baryons in virialized systems.  
On the whole, our results imply a fairly intuitive picture of the influence of 
$\Omega_b$ on high-redshift structure, and we provide scalings that can be used
to estimate the impact of $\Omega_b$ uncertainties on the predictions of 
hydrodynamic simulations.

\end{abstract}

\section{Introduction}

In the past half decade, estimates of the cosmic baryon density
parameter $\Omega_b$ have ranged from as low as $0.005h^{-2}$ to as
high as $0.030h^{-2}$, driven mainly by studies of the deuterium
abundance in high-redshift Ly$\alpha$ absorbers and by measurements of
anisotropy in the cosmic microwave background (CMB).  The primordial
deuterium abundance, combined with the theory of big bang
nucleosynthesis, should in principle yield a tight constraint on the
quantity $\Omega_b h^2$.  However, estimates of the deuterium
abundance in Ly$\alpha$ absorbers have ranged widely \citep[e.g.][]
{songaila94,rh96,webb97,burles98ab,PB01,dodorico01,omeara01}, 
with a recent consensus
emerging in favor of $D/H\approx 2-3\times 10^{-5}$ and a
corresponding $\Omega_b h^2 \approx 0.02-0.025$.  The first analyses
of the BOOMERANG and MAXIMA experiments implied a low second peak in
the CMB power spectrum, which could be explained by a higher baryon
density $\Omega_b h^2 \approx 0.03$ \citep{boomerang,maxima,PS00}.
Results from the more recent analysis of BOOMERANG (including more of
the data) and from the DASI experiment show a stronger second peak
that is consistent with the $\Omega_b h^2$ values inferred from the
deuterium abundance \citep{netterfield01,pryke01}.  It is tempting to
conclude, therefore, that the baryon density is now known to be
$\Omega_b = 0.022 h^{-2}$ with an uncertainty $\sim 10\%$.  However,
recent history suggests that we should remain somewhat cautious
about this conclusion until it has stood for a longer period of time,
especially since there are still discrepancies between this value of
$\Omega_b h^2$ and most estimates of the $^4$He abundance within the
framework of standard big bang nucleosynthesis \citep{tytler00,KS00}.

Given the remaining uncertainties, it is important to understand the
impact of $\Omega_b$ on the predicted properties of cosmic structure.
We investigate this issue using hydrodynamic
simulations that have identical cosmological and numerical parameters
and differ only in the value of $\Omega_b$.  Even if one believes that
the value of $\Omega_b$ is well constrained by observations, an
investigation that isolates the effect of the baryon density can give
insight into the physics that governs the processes of galaxy
formation and the state of the intergalactic medium.

We have previously used simulations like the ones carried out here to
predict the evolution of different phases of the intergalactic medium
(Dav\'e et al. 1999), the properties of the Ly$\alpha$ forest (Hernquist
et al.\ 1996), the properties of damped Ly$\alpha$ and Lyman Limit
systems (Gardner et al.\ 2001, hereafter GKHW, and references therein), 
and the masses
and star formation rates of high-redshift galaxies (Weinberg, Katz, \&
Hernquist 2001).  These are the quantitative predictions that we focus
on here, restricting our attention to high-redshift structure purely
for reasons of computational practicality: it takes much less computer
time to evolve three simulations to $z=2$ than to $z=0$.  
Although the qualitative effects of changing $\Omega_b$ are usually easy 
to guess, the scaling of quantities with $\Omega_b$ is not obvious and 
turns out in some cases to be non-intuitive.  These scalings give guidance
to the theoretical uncertainties associated with uncertainties in $\Omega_b$
and a better understanding of the role of gas physics and radiative
processes in determining the properties of high-redshift structure.

\section{Simulation and Methods}

In this paper we present the results of three simulations of the
``standard'' cold dark matter model ($\Omega_m=1$, $h \equiv H_0/100$
\kmpersec\ Mpc$^{-1}=0.5$, $\sigma_8=0.7$) identical in every respect
except the baryonic mass fraction $\Omega_b$, which is set to $0.02,
0.05$ and $0.125$.  All three runs are performed in a manner similar
to that described by GKHW and Katz, Weinberg, \& Hernquist (1999), wherein 
a periodic
cube whose edges measure 11.11\hinv Mpc in comoving units is drawn
randomly from a CDM universe and evolved to a redshift $z=2$.  The
simulations employ $64^3$ gas and $64^3$ dark matter particles, with a
gravitational softening length of 5\hinv\ comoving kpc (3\hinv\
comoving kpc equivalent Plummer softening, $1h^{-1}$ physical kpc at
$z=2$).  The dark matter particle mass is $2.8 \times 10^9 M_\odot$
and the gas particle masses are $5.8 \times 10^7$, $1.5 \times 10^8$,
and $3.6 \times 10^8 M_\odot$ for $\Omega_b= 0.02, 0.05$, and 0.125
respectively. This yields baryonic mass resolutions (defined by a 64-particle
threshold) of $3.7 \times 10^9$, $9.6\times 10^9$, and $2.3 \times 10^{10}
M_\odot$ for the respective $\Omega_b$'s.

Detailed descriptions of the simulation code and the radiation physics
can be found in Hernquist \& Katz (1989) and Katz, Weinberg, \&
Hernquist (1996; hereafter KWH), and we only summarize the techniques
here.  We perform our simulations using \tsph\ (\citealt{hk89}), a code
that unites smoothed particle hydrodynamics (SPH; \citealt{lucy77};
\citealt{gingold77}) with a hierarchical tree method for computing
gravitational forces (\citealt{bh86}; \citealt{h87}).  Dark matter, stars,
and gas are all represented by particles; collisionless material is
influenced only by gravity, while gas is subject to gravitational
forces, pressure gradients, and shocks.  We include the effects of
radiative cooling, assuming primordial abundances, and Compton
cooling.  Ionization and heat input from a UV radiation background are
incorporated in the simulation.  We adopt the UV background spectrum
of Haardt \& Madau (1996), although it is reduced in intensity by a
factor of two at all redshifts for the $\Omega_b = 0.02$ and 0.05 simulations so
that the mean flux decrement of the \lya\ forest (LAF hereafter) 
is closer to the 
observed value given our assumed baryon density (\citealt{croft97}).  We apply
further adjustments to the background intensity during the
analysis stage to match the Press, Rybicki, \& Schneider
(1993) measurements of the mean decrement, as discussed further 
in \S\ref{subsec:laf} below.  We use a
simple prescription to turn cold, dense gas into collisionless
``star'' particles and return the resulting supernova feedback energy to the
surrounding medium as heat.  The prescription and its computational
implementation are described in detail by KWH.  Details of the
numerical parameters can be found in Katz et al. (1999).

\subsection{Halo and Galaxy Identification}
\label{ssec:ident}

>From the simulation outputs at $z=4,$ 3, and 2, we identify dark
matter halos and the individual concentrations of cold, collapsed gas
that they contain.  We identify the halos by applying a
friends-of-friends algorithm (FOF) to the combined distribution of
dark matter and SPH particles, with a linking length equal to the mean
interparticle separation on an isodensity contour of an isothermal
sphere with an enclosed average overdensity of 178, the virial
overdensity.  In an isothermal sphere, the local density at the virial
radius is simply one third the virial overdensity.  The dark matter particles in
FOF-identified groups correspond to the dark matter halo of a galaxy
or cluster.  For all of the analyses in this paper, we impose a cutoff
of 64 dark matter particles, corresponding to a halo mass
$M_{res}=1.9\times 10^{11}M_\odot$, to eliminate incompleteness effects
at smaller, marginally resolved masses. 

To detect galaxies within the dark matter halos, we search for
discrete regions of cold collapsed gas and stars (CCGS) by applying
the algorithm Spline Kernel Interpolative DENMAX (``SKID'') of Stadel \ea  
(2000; see also KWH and
http://www-hpcc.astro.washington.edu/tools/skid.html) to the distribution
of baryonic particles.  SKID identifies gravitationally bound groups
of gas and star particles that are associated with a common density
maximum.  Gas particles are only considered as potential members of a
SKID group if they have temperature $T < 30,000$ K and smoothed
density $\rho_g/\bar\rho_g > 1000$.  We discard SKID groups with fewer
than eight members.  As in GKHW, we find that all of the gas
concentrations found by this method reside within the larger dark
matter halos identified by FOF.  The particles identified to be part
of the remaining SKID groups are labeled CCGS (Cold Collapsed Gas and Stars),
and we consider each SKID group
to be a galaxy.  The total CCGS mass of a halo is calculated
by summing the mass of particles in SKID groups within a given FOF group.

\section{Results}

We present results spanning a wide range of baryonic temperatures and densities.
First we discuss global thermal and density distributions of the baryonic
matter, including implications for the Ly$\alpha$ forest and
the mean opacity of the Universe.  We then concentrate on baryons
resident within collapsed objects, examining the effects of $\Omega_b$
on damped Ly$\alpha$ absorber and Lyman limit incidences, halo baryon and 
CCGS fraction, and star formation.

\subsection{Global Baryonic Behavior}
\label{ssec:global}

\begin{figure}
\vglue-0.65in
\plottwo{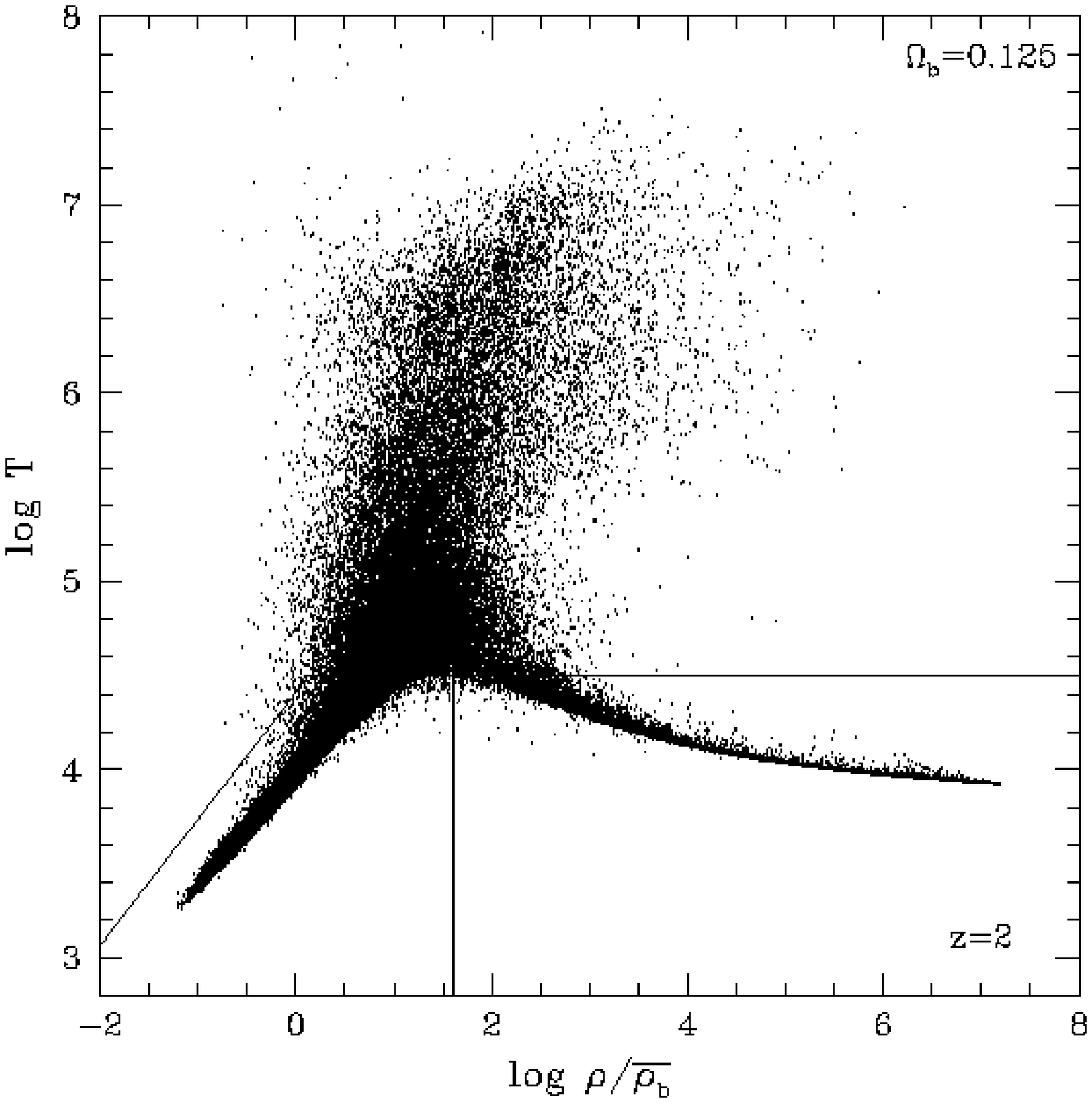}{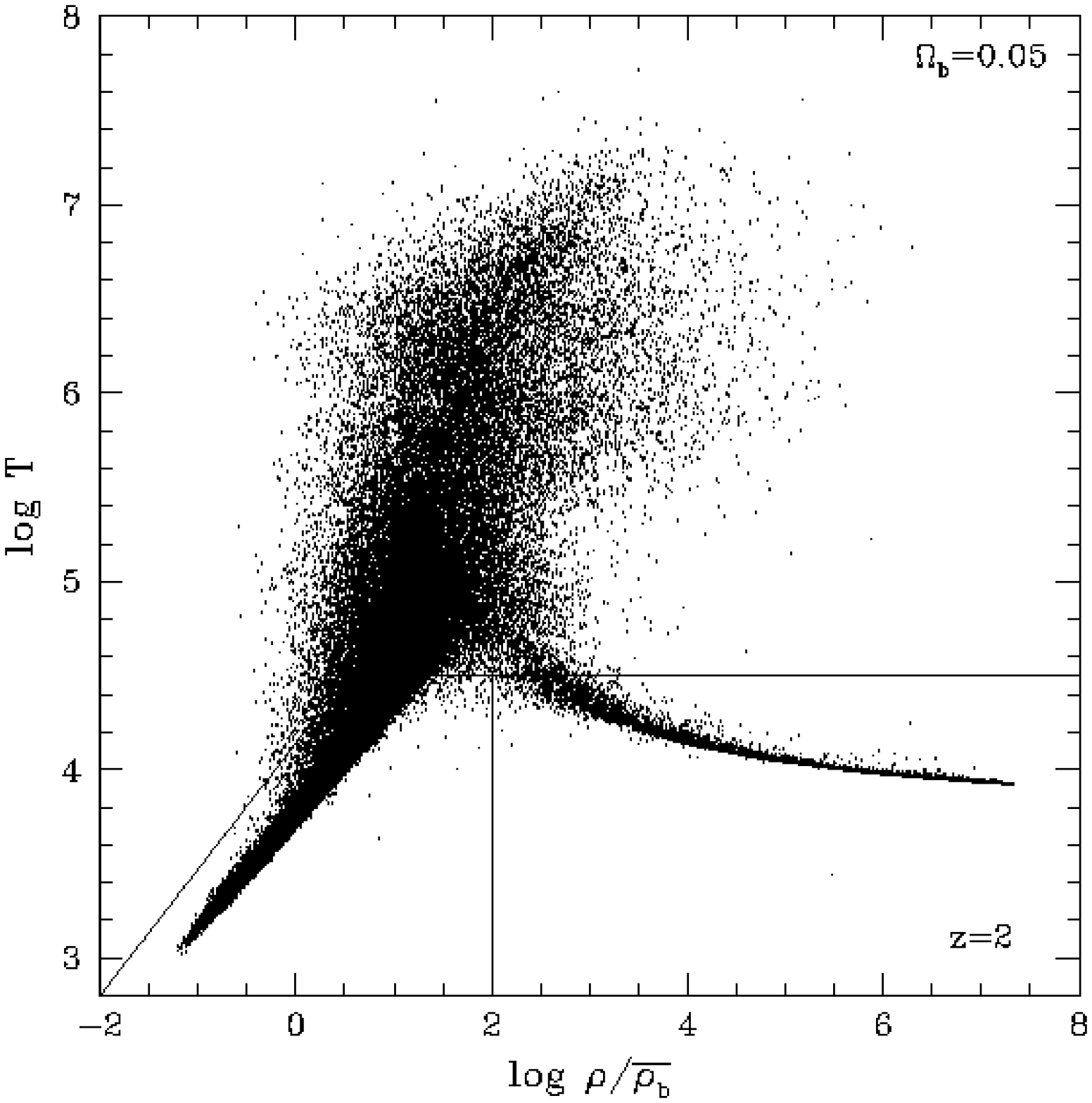} \\
\columnwidth=2.22222\columnwidth 
\vglue-0.05in
\plotonesmall{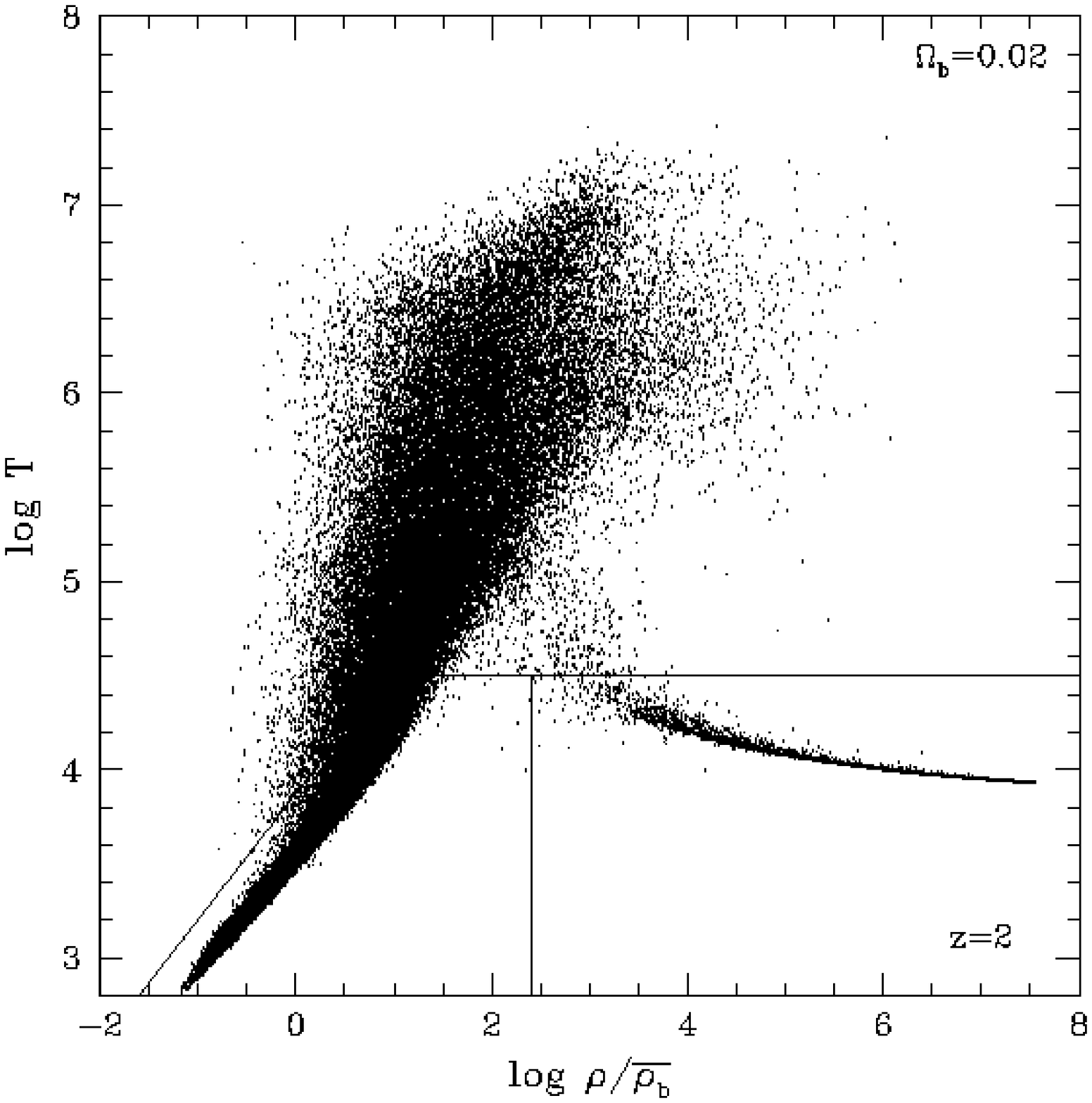} \\
\vglue-0.05in
\caption{Gas particle density $\rho$ in units of the mean baryon
density $\bar{\rho}_b$ vs.\ temperature $T$.  The solid lines indicate the
borders for gas particles to be considered cool, photoionized, diffuse
IGM (lower left box); condensed gas (lower right box); and
shocked IGM (upper remainder of panel).}
\label{fig:rhot}
\end{figure}

Figure~\ref{fig:rhot} shows the temperature-density distribution for
$10^5$ randomly selected gas particles in each simulation.  We divide
$\rho-T$ space into the three major populations according to the
prescription in Dav\'e \ea (1999), using the same cutoffs in
temperature and physical density:
\vglue-0.2in
\begin{enumerate}
\itemsep=0pt
\parsep=0pt
\item {\it Condensed:} ($\log T < 4.5$ and $\log\rho/\bar{\rho}_b
> 2$) cold, dense gas associated with galaxies.
\item {\it Diffuse:} ($\log T < 4.5$ and $\log\rho/\bar{\rho}_b <
2$ and $\log T < 2/3(\log\rho/\bar{\rho}_b+2)+2.8$) cool, ionized,
diffuse IGM heated by photoionization but able to cool adiabatically.
\item {\it Shocked:} (remainder) shocked IGM in the form of
shock-heated gas in filaments and halos.
\end{enumerate}

We adopt the term ``condensed'' to denote gas classified simply by
this $\rho-T$ cut to differentiate it from ``cold collapsed gas'' ---
part of the cold collapsed gas and stars (CCGS) component presented
later on --- which is gas that is part of a SKID-identified group (and
hence gravitationally self-bound).

\begin{figure}
\vglue-0.65in
\plotone{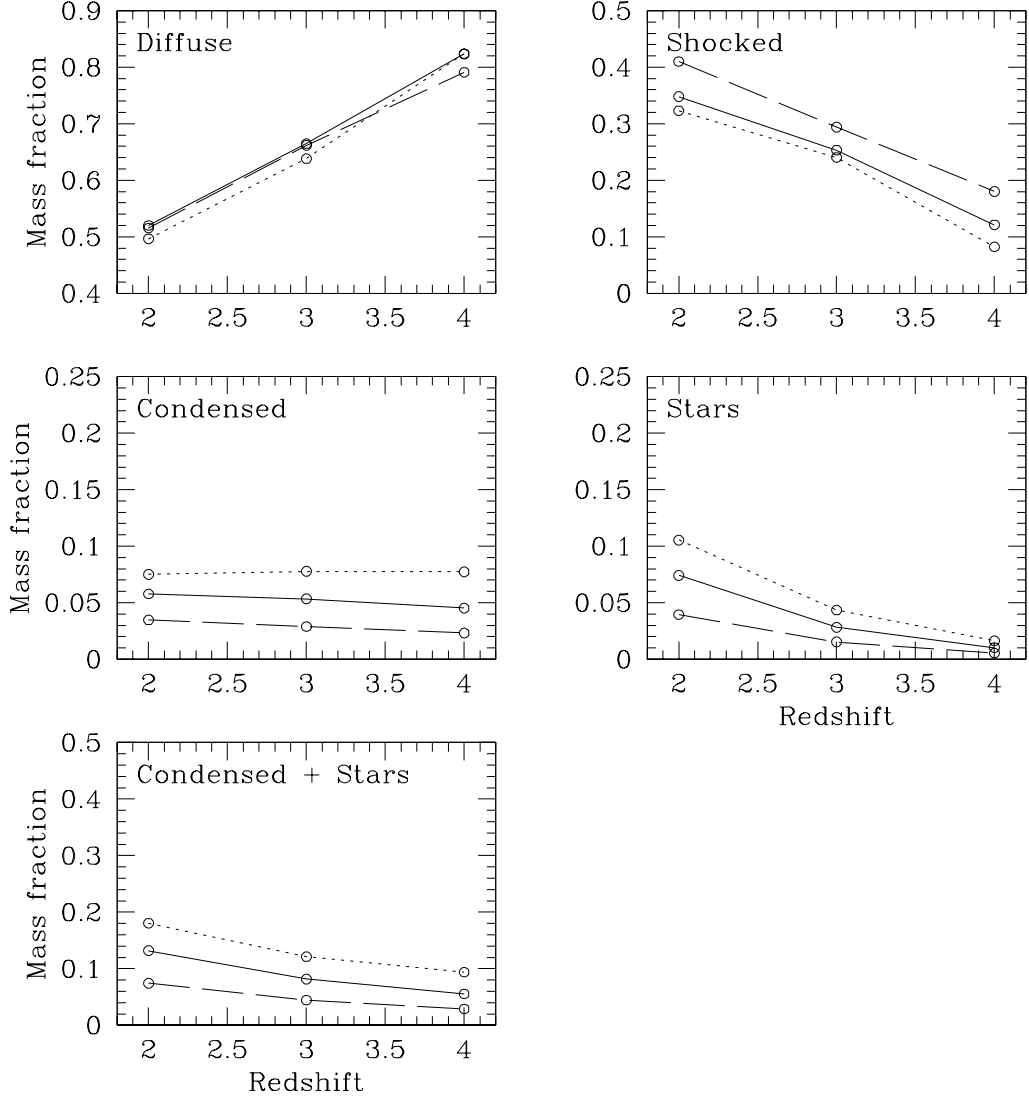} \\
\vglue-0.2in
\caption{ Evolution of the mass in the various phases in units of the
total baryon mass.  The three principal gas phases --- diffuse,
shocked, and condensed --- are given in the first three panels.  The
fourth panel shows the stellar mass fraction, and the bottommost panel shows
condensed gas + stars.  The line styles are consistent throughout the
paper and are as follows: $\Omega_b=0.125$ dotted, $\Omega_b=0.05$
solid, and $\Omega_b=0.02$ long-dashed.  
Note that the $y$-axes of the plots differ slightly to 
enhance differences between models.  
}
\label{fig:phases}
\end{figure}

Figure~\ref{fig:phases} compares the mass fraction in each of the
three gaseous phases from Figure~\ref{fig:rhot} as a function of redshift.
Stellar mass
fraction is displayed separately and in combination with condensed
gas.  Hence, the bottom-left panel is the true measure of condensed
matter in the simulation.  The line styles denoting the different
simulations are the same throughout this paper: $\Omega_b=0.125$
dotted, $\Omega_b=0.05$ solid, and $\Omega_b=0.02$ long-dashed.  The
higher gas density in high $\Omega_b$ cosmologies clearly promotes
more aggressive cooling, which in turn enhances star formation
activity.  The fraction in the diffuse component reflects the original state of the
gas and is largely unaffected by variations in global baryon density.
Hence, the increased mass in condensed baryons is offset by a deficit
in shocked gas.  The fact that the condensed and shocked phases
establish a balance while leaving the diffuse component unchanged
indicates that the gas is shocked in high-density regions physically
proximate to the cooling gas, but not in regions that harbor the
diffuse component.

Table~\ref{tab:fits} shows the results of fitting a power law in
$\Omega_b$ for each of the five components.  As one would surmise from
Figure~\ref{fig:phases}, the diffuse component exhibits little
$\Omega_b$ dependence.  The baryon dependence of the shocked component
is difficult to characterize by a power-law because it varies
more strongly when changing $\Omega_b$ from 0.02 to 0.05 than from
0.05 to 0.125; in the latter case, the removal of shocked gas to 
the condensed/stellar phases is partly compensated by a small drop
in the diffuse phase.
The fractions of baryons in the condensed and stellar phases
scale fairly well with $\Omega_b$ and can be
approximated by power laws that vary approximately as
$\Omega_b^{1.5}$, with the exact scaling depending on the precise phase and
redshift.  The errors in the $\chi^2$ fit to the exponent are minimal,
although it is difficult to assign formal errors to the quantities in
Figure~\ref{fig:phases}.

\begin{table}
\begin{tabular}{llllr}
\tableline\tableline
\multicolumn{1}{c}{ }&\multicolumn{3}{c}{$\beta$}&
\multicolumn{1}{c}{$\beta(z)$} \\ 
 & $z=2$&$z=3$&$z=4$&  \\ \cline{2-5}
Diffuse		&0.980  &0.980   &1.022  &\multicolumn{1}{l}{minimal dependence} \\
Shocked		&0.870  &0.890   &0.572  &\multicolumn{1}{l}{no simple dependence} \\
Condensed	&1.419  &1.539   &1.656  &$0.12 (z-3) + 1.54$ \\
Stars		&1.534  &1.568   &1.596  &$0.031 (z-3) + 1.56$ \\
Condensed+Stars &1.484  &1.549   &1.645  &$0.081 (z-3) + 1.56$ \\
 & $z=2$&$z=4$&$z=6$&  \\ \cline{2-5}
Star Formation Rate &1.434 &1.561&1.714  &$0.070 (z-3) + 1.50$ \\
\tableline\tableline
\end{tabular}
\caption{ Table of baryonic dependence of global baryon components and other
global quantities.  The dependence is given as a power law $\Omega_b^\beta$
shown in columns 2-4.  The last column shows the best fit linear
dependence of $\beta$ on redshift, centered on redshift $z=3$.  The first
five lines give the scaling of the mass in the components given in
Figure~\protect\ref{fig:phases}.  
The last line gives the $\Omega_b$ scaling of the 
global star formation rate
(Figure~\protect\ref{fig:madauplot}).}
\label{tab:fits}
\end{table}

\subsection{The Lyman-alpha Forest}
\label{subsec:laf}
\begin{figure}
\vglue-0.65in
\plottwo{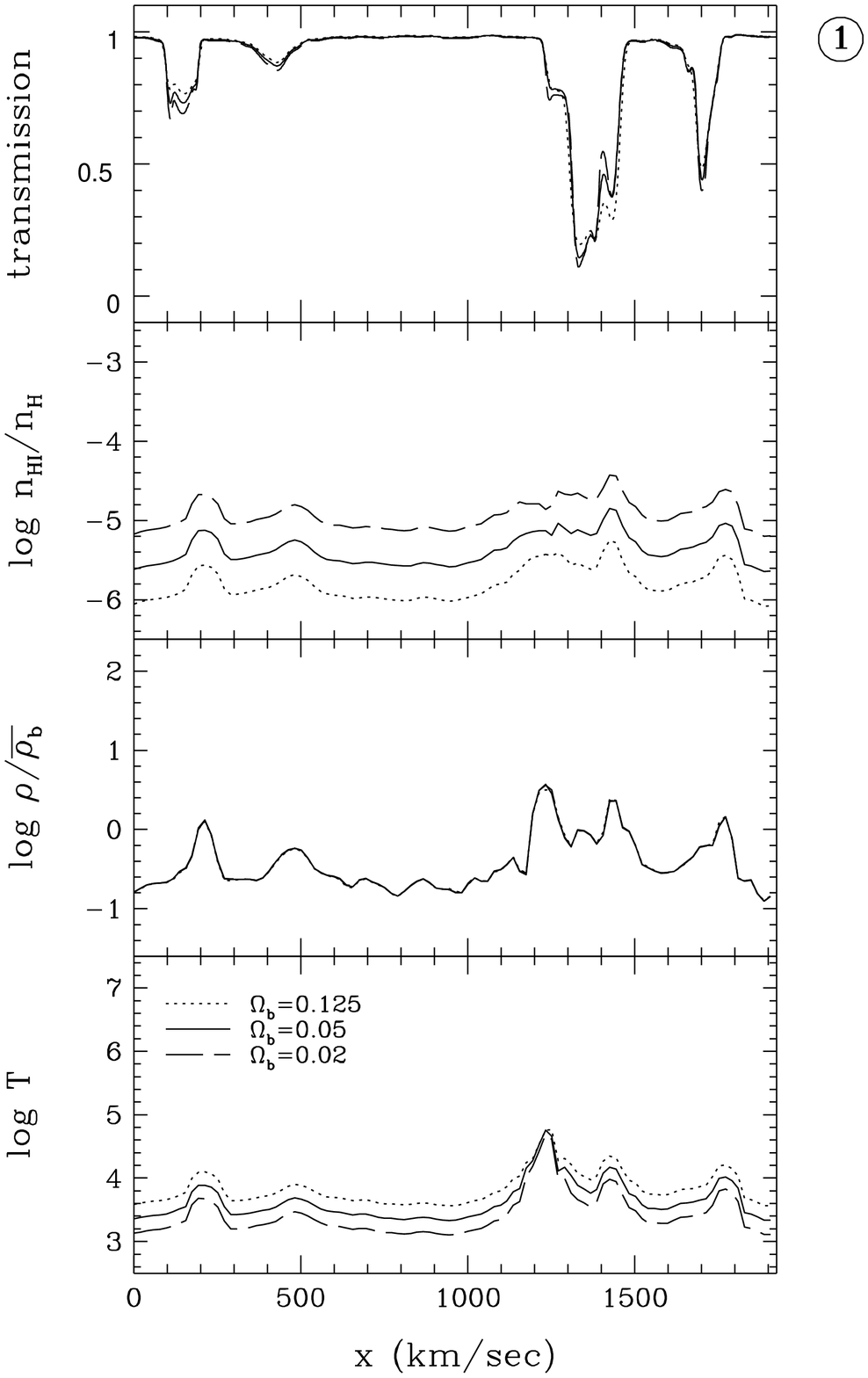}{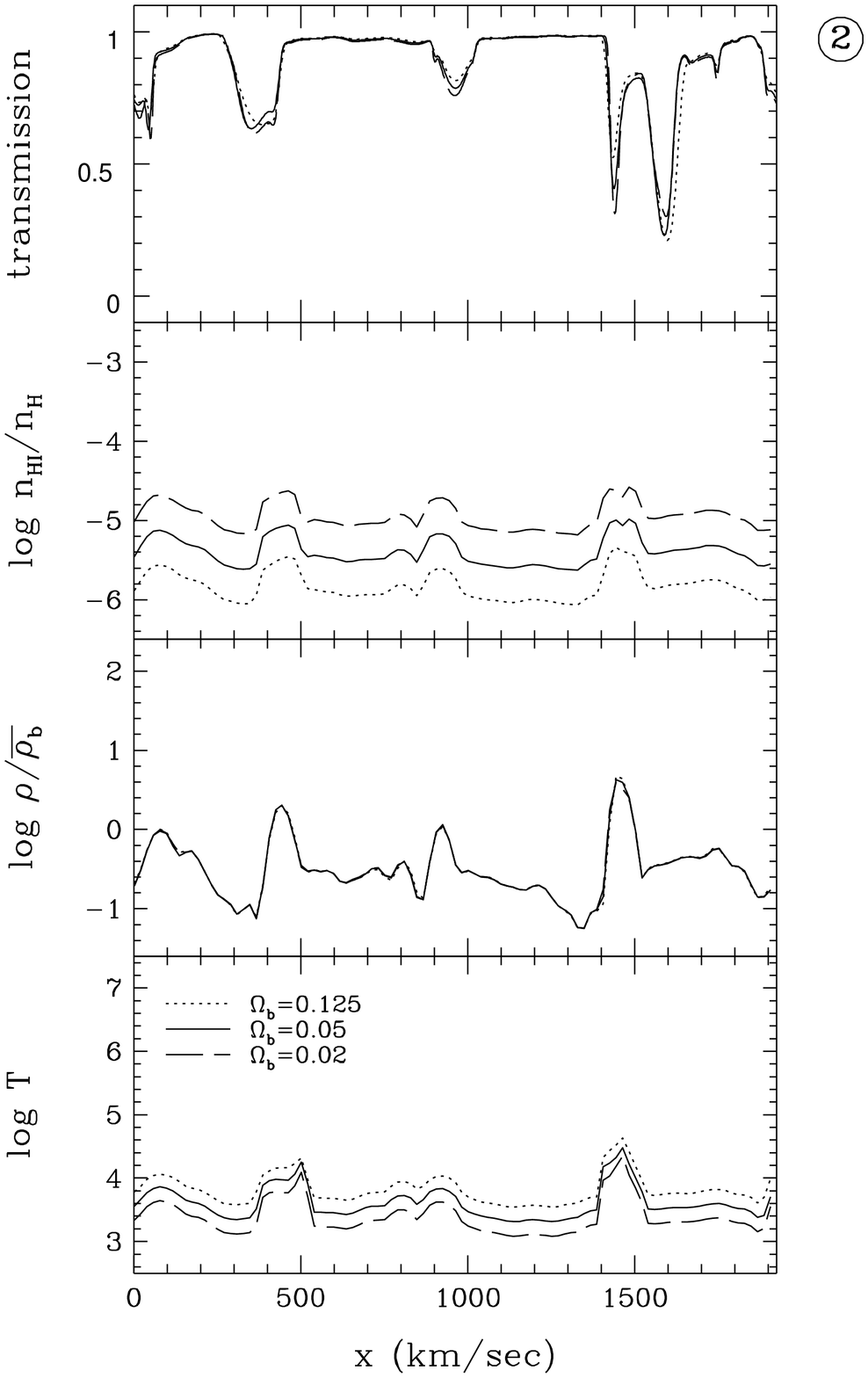} \\
\columnwidth=2.22222\columnwidth 
\vglue-0.2in
\plottwo{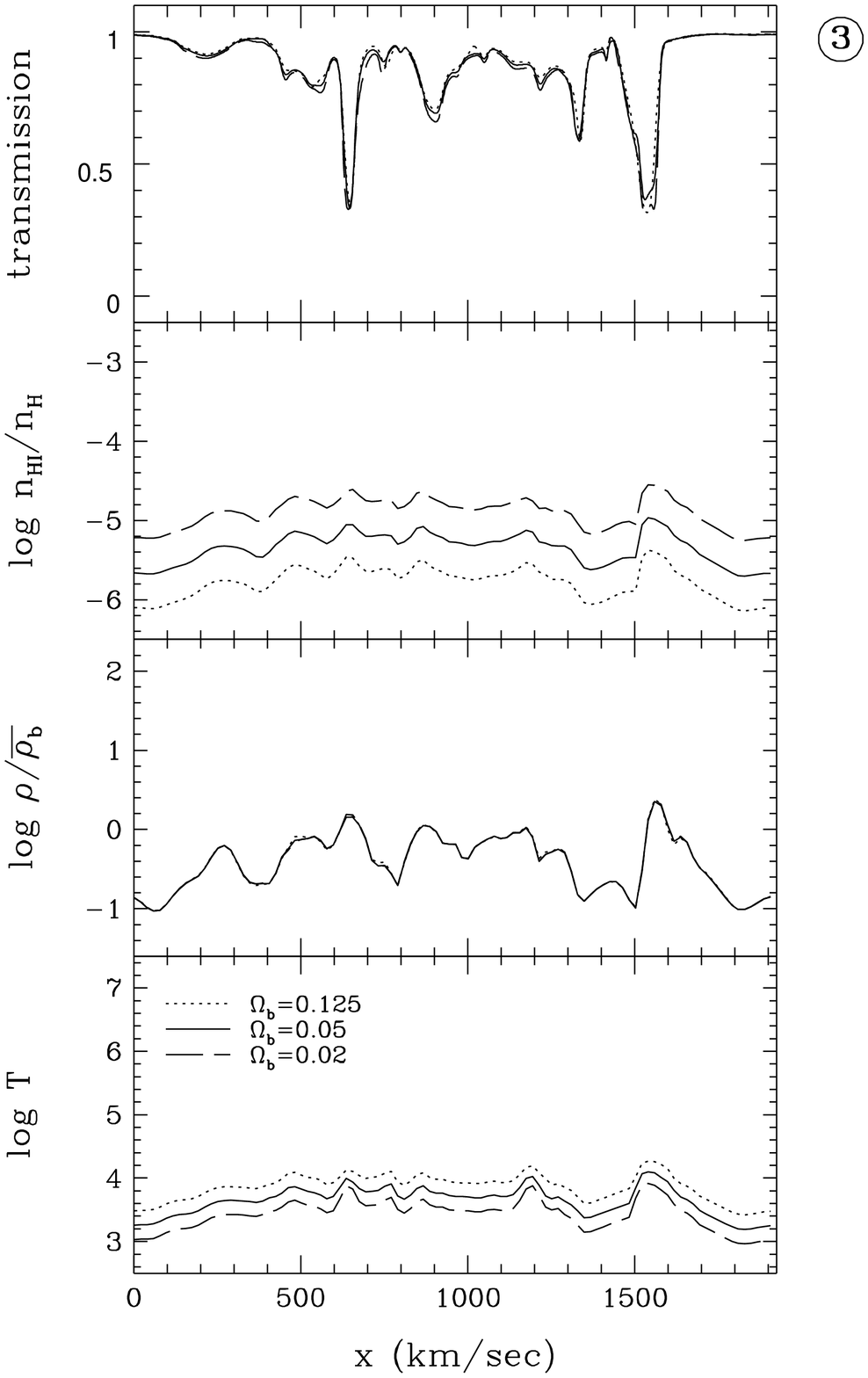}{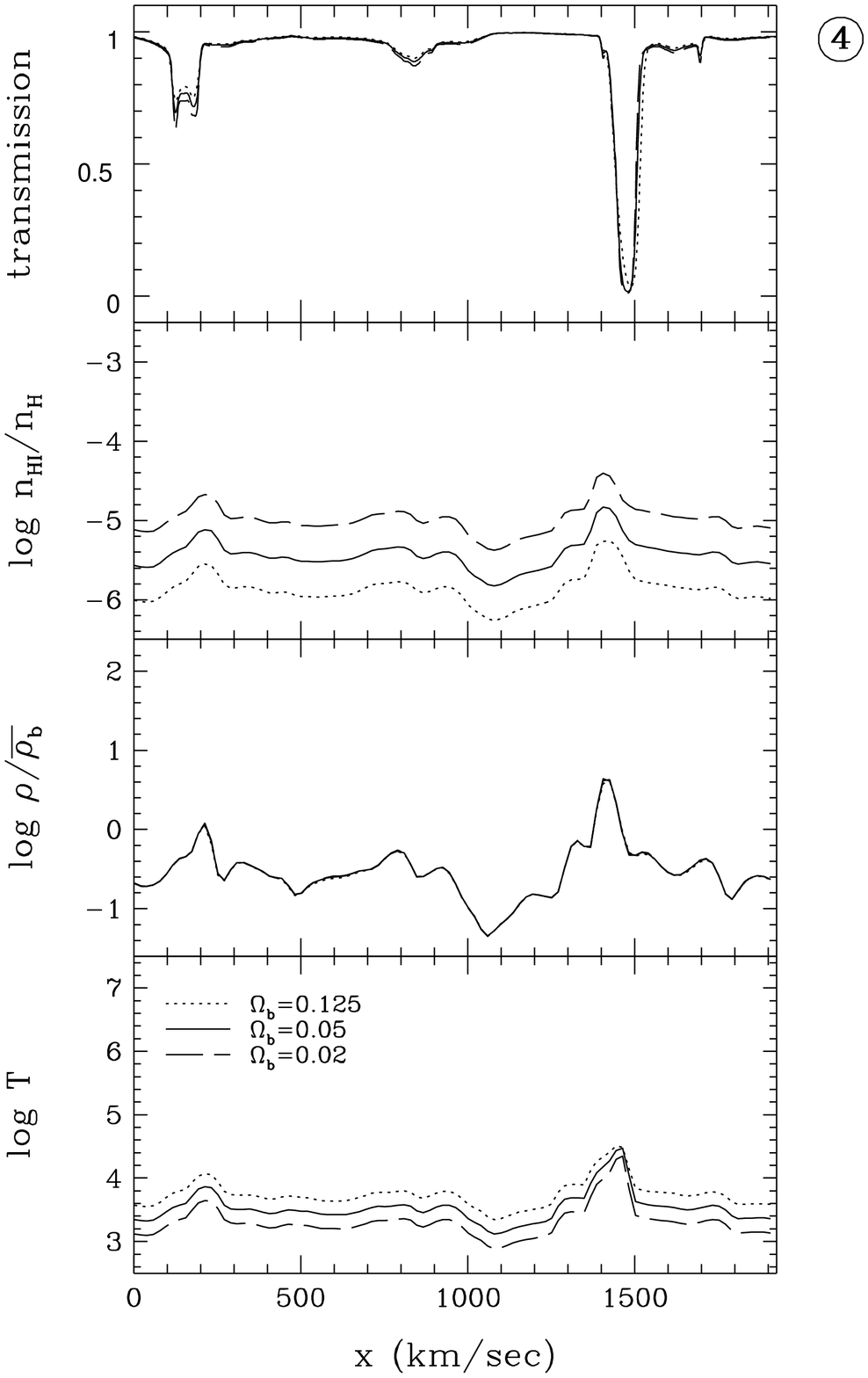} \\
\vglue-0.2in
\caption{Profiles of (top to bottom) transmitted flux, hydrogen neutral
fraction, gas overdensity, and temperature, along four lines of sight
through each of the three simulations at $z=2$.  In each simulation, the
intensity of the UV background is adjusted so that the simulation reproduces
the observed mean decrement of the LAF.  As a result, higher neutral
fractions compensate for the lower gas densities in the lower $\Omega_b$ 
models, and the three simulations yield nearly identical LAF spectra.}
\label{fig:zooplots}
\end{figure}

To study the influence of $\Omega_b$ on LAF absorption, we create
synthetic spectra using the TIPSY package \citep{TIPSY}, following
the algorithm described by \citet{HK96}.
If we used the same UV background intensity in all three simulations,
then the higher $\Omega_b$ runs would have higher gas densities,
higher neutral fractions, and a higher mean opacity of the LAF.
Instead we follow the by now standard procedure of adjusting
the UV background intensity (which is observationally quite uncertain)
so that each simulation reproduces the observed mean opacity
(see, e.g., \citealt{miralda96,croft97}).  Table~\ref{tab:mean}
lists the factors by which we must multiply the \citet{HM96}
background intensities to match the Press et al.\ (1993) values of
the LAF mean flux decrement in the three simulations at each redshift.
These factors scale roughly as $\Omega_b^{1.7}$.

Figure~\ref{fig:zooplots} shows spectra and 1-d profiles along four
randomly selected lines of sight through the simulations at $z=2$.
The gas density profiles (2nd panel up in each case) are virtually
identical when computed in units of the mean baryon density.
Random lines of sight intersect regions of low to moderate overdensity,
where changing the baryon fraction has negligible impact on the
density field.  Neutral fractions (3rd panel up) are higher for
lower $\Omega_b$, as they must be to maintain a constant value of 
the mean flux decrement
(which scales roughly with the mean neutral hydrogen density).
Temperatures (bottom panel) are higher for higher $\Omega_b$ because 
the higher photoionization rate leads to more rapid heating, while
the cooling, which is dominated by adiabatic expansion in these low
density regions, stays constant.  These higher temperatures
are the reason that the required UV background intensity scales
as $\Omega_b^{1.7}$ rather than $\Omega_b^2$ as one might naively expect.
Higher temperatures reduce the recombination coefficient, so the
neutral fraction at fixed photoionization rate does not grow as
fast as $\Omega_b$.

The top panels in Figure~\ref{fig:zooplots} show the synthetic 
spectra themselves, and they are virtually indistinguishable 
for the three different $\Omega_b$ values.  Because the gas overdensity
structure is nearly independent of $\Omega_b$, normalizing the UV
background to match a fixed mean decrement makes the spectra match
point by point, with only a minor difference caused by differences
in thermal broadening.  Figure~\ref{fig:zooplots} thus provides 
strong backing for the standard practice of scaling LAF spectra 
derived from hydrodynamic simulations.  When gas temperatures are
set by the interplay between photoionization heating and adiabatic
cooling, as they are here, then the parameter combination constrained
by the observed mean opacity scales as $\Gamma/\Omega_b^{1.7}$, where
$\Gamma$ is the HI photoionization rate.

\begin{table}
\begin{tabular}{lrrr}
\tableline\tableline
$\Omega_b$ &\multicolumn{1}{c}{$z=2$}  &\multicolumn{1}{c}{$z=4$}
&\multicolumn{1}{c}{$z=4$} \\
\tableline
0.125	& 1.90  & 1.78	& 5.34 \\
0.05	& 0.40	& 0.40	& 1.20 \\
0.02	& 0.08	& 0.09	& 0.27 \\
\tableline
\end{tabular}
\caption{ Table of factors by which the Haardt \& Madau
(1996) UV background intensity must be multiplied in order for the
mean opacity of the simulation volume to match Press et al.\ (1993) values.}
\label{tab:mean}
\end{table}

\subsection{Damped Ly$\alpha$ and Lyman Limit Systems}

\begin{figure}
\vglue-0.65in
\plottwo{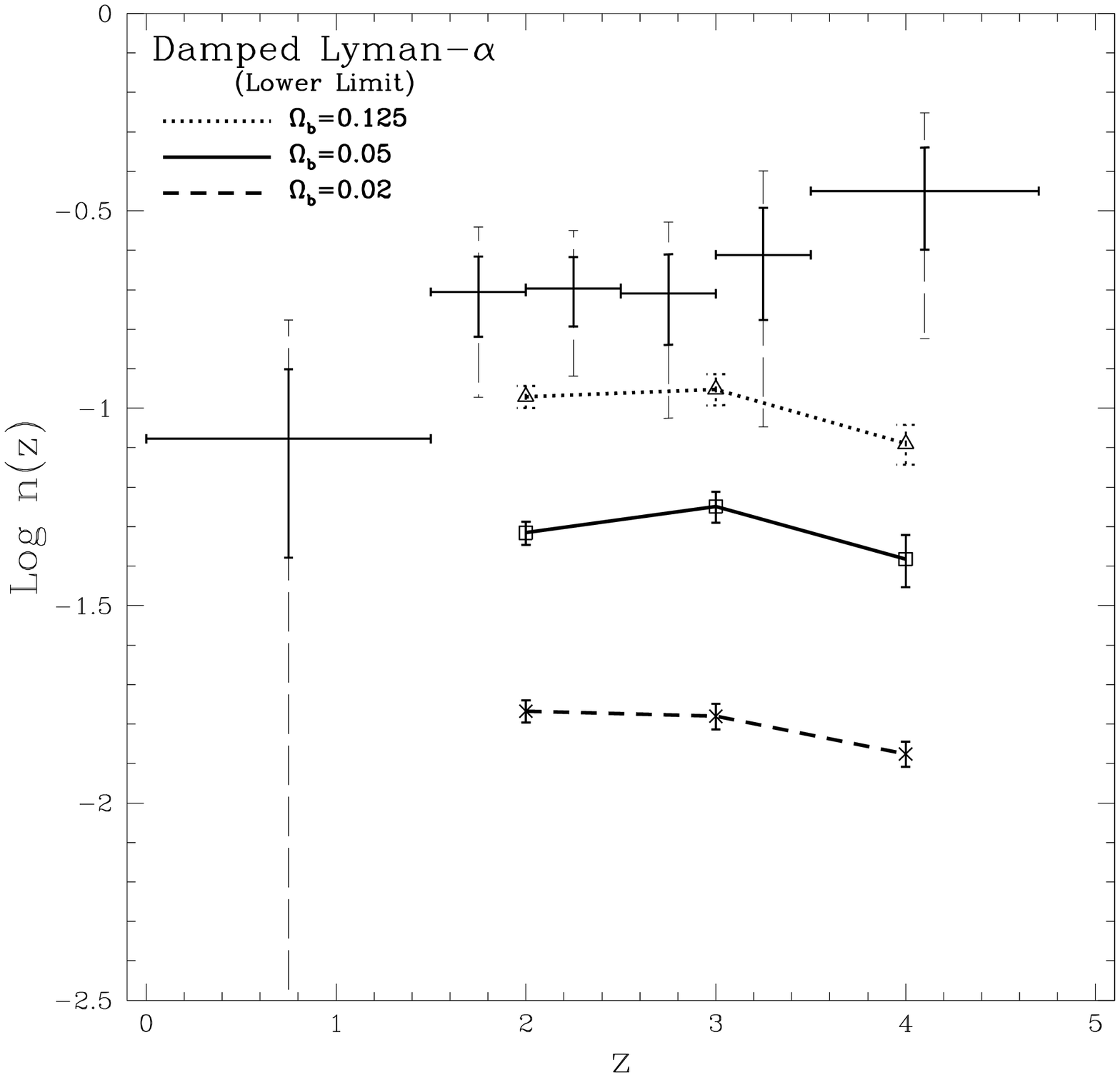}{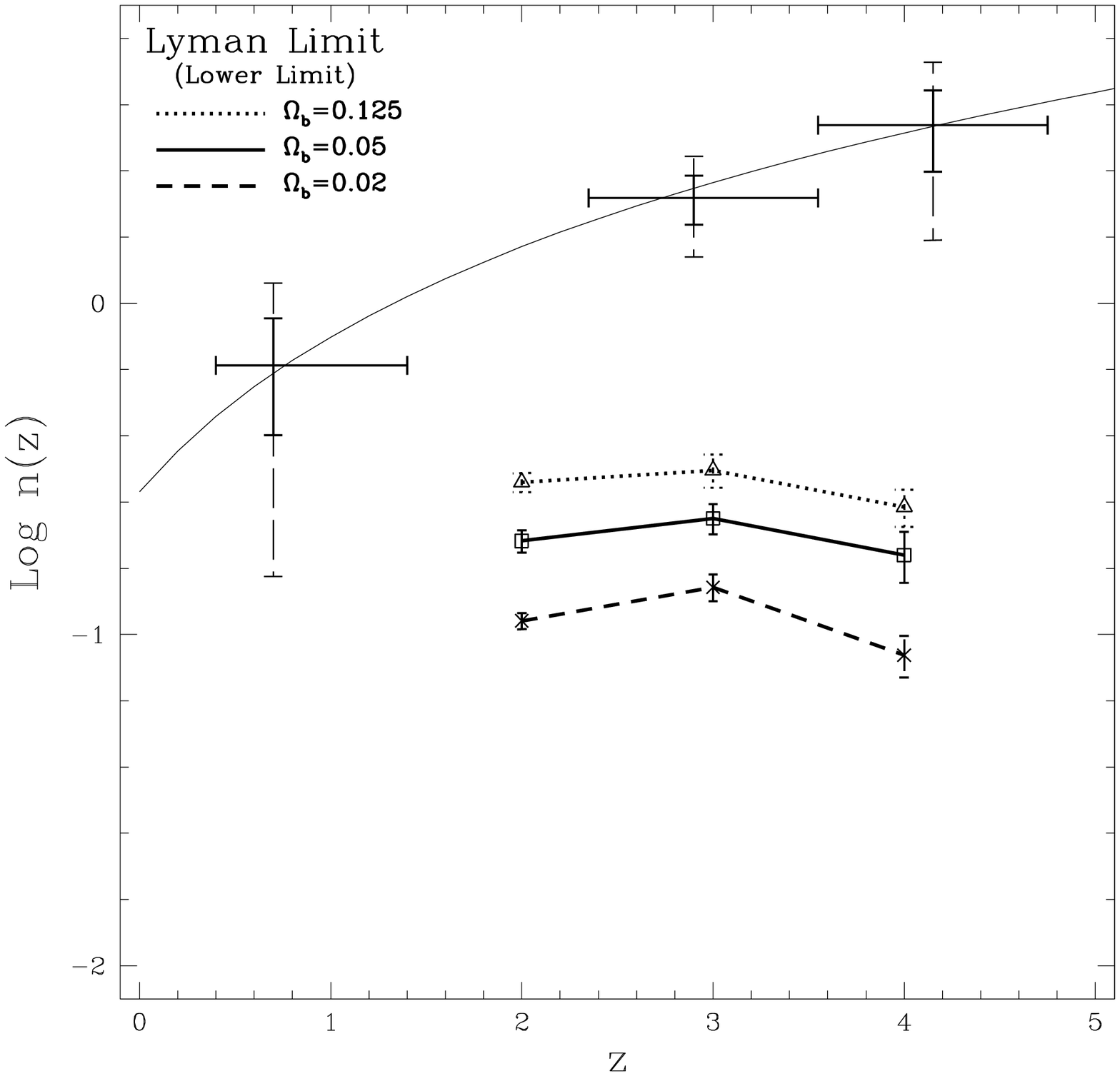} \\
\vglue-0.2in
\caption{ Effects of $\Omega_b$ on damped Ly$\alpha$ absorber and Lyman limit
incidence $n(z)$.  
We evaluate all models at the same
redshifts $z=2,3,4$ but plot them with slightly offset $z$ values for
greater clarity.  
Because the simulations only resolve absorbers in halos with
$v_c \geq 100$ \kmpersec, they are expected to underestimate 
the true incidence of absorption, perhaps by a large factor
(see GKHW).
These are plotted against damped Ly$\alpha$ absorber
observational values
from \protect\cite{slw00} and Lyman limit observational
values from Storrie-Lombardi \ea (1994).  The observations are shown
with $1\sigma$ (solid) and $2\sigma$ (dashed) error bars.  
Error bars on the simulation points denote the 68.3\% confidence limit.}
\label{fig:DLALL}
\end{figure}

We measure the incidence $n(z)$ of damped Ly$\alpha$ (DLA) and Lyman
limit (LL) systems in each simulation using the method described in
GKHW.  By ``incidence'' we mean the number of systems with $N_{\rm
HI}\geq 10^{20.3}\;\cdunits$ (DLA) or $N_{\rm HI} \geq
10^{17.2}\;\cdunits$ (LL) intercepted along a line of sight per unit
redshift.  Although our simulations only resolve halos with virial
radius circular velocities $v_c \simgt 100$ \kmpersec, GKHW find that
a significant fraction, possibly even the majority, of the DLA and LL
incidence arises in halos less than this cutoff.  Consequently, in
comparing our simulated $n(z)$ to observational values in
Figure~\ref{fig:DLALL}, we would expect our results to be below what
is observed and serve as lower limits.  However, given that all three
simulations have the same mass resolution, we can compare the
simulations against one another to understand how DLA and LL incidence
is affected by $\Omega_b$.  

GKHW and Gardner \ea (1997) demonstrated that the cross-section of
halos harboring DLA and LL absorbers was a complex function of halo
mass, interaction history, and cooling time.  Specifically, absorbers
that are able to cool more quickly tend to have more mass in cooled
neutral gas but also {\em smaller} cross-sections as the gas tends to
collapse more tightly.  The competing effect that helps determine the
halo absorption cross-section is that higher mass halos tend to harbor
more absorbers, thus increasing the overall absorption cross-section
with mass.  The incidence of DLA and LL systems, however, follows the
intuitive expectation that a higher $\Omega_b$ means more absorbers.
If we examine the results on a halo by halo basis (not shown) we find
that, in general, the scaling in Figure~\ref{fig:DLALL} results from
halos of all masses having uniformly larger absorption
cross-sections in higher $\Omega_b$ models.  Hence, although the
complex link between gas mass and DLA/LL cross-section need not have
yielded an intuitive relationship between cross-section and
$\Omega_b$, DLA and LL incidence does appear to scale in a relatively
simple manner with the universal baryon fraction.

\subsection{Baryonic Behavior Within Dark Matter Halos}

\begin{figure}
\vglue-0.65in
\plotone{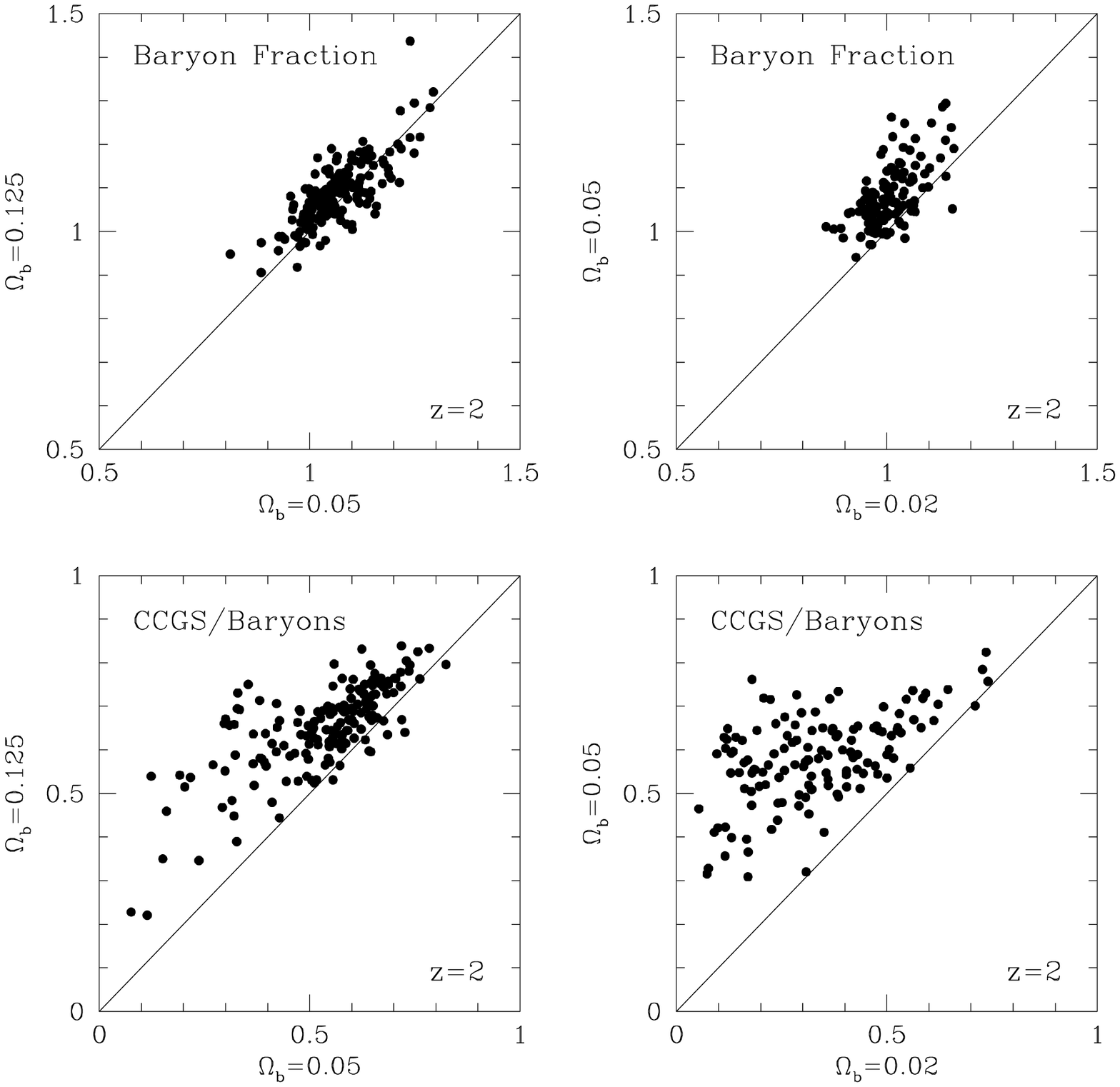} \\
\vglue-0.2in
\caption{
Effect of $\Omega_b$ on halo baryon fraction and cooling fraction at
redshift $z=2$.
Each data point
represents a specific halo which was identified in the two
simulations labeled on the $x$ and $y$ axes.  The upper panels compare
the baryon fraction of a FOF-identified halo in each simulation, normalized to
the Universal value, $\Omega_b/\Omega_m$.  The lower panels plot the fraction of
cold collapsed gas and stars (CCGS) with respect to the total baryonic
content of the halo.  The solid line indicates the locus of no dependence on cosmic
baryon abundance.}
\label{fig:baryonfrac}
\end{figure}

We now examine the behavior of baryons within collapsed objects,
specifically groups which have been identified with the FOF algorithm
and are above the mass resolution cutoff.  Cold collapsed gas and stars
(CCGS) in each halo are identified using SKID as detailed in
section~\ref{ssec:ident}.  Figure~\ref{fig:baryonfrac} compares the
baryonic fraction and the fraction of baryons in CCGS in each of the
simulations.  Given that the simulations have the same total mass and
were run from exactly the same initial waves, it is simple to identify
the corresponding halos in each simulation.  Each point in
Figure~\ref{fig:baryonfrac} represents the same halo 
identified in all three simulations and matched based on its position.
Looking at the
top panels, we notice that at $z=2$ the baryonic fraction of halos
normalized to the baryonic fraction of the Universe as a whole is insensitive
to $\Omega_b$ for $\Omega_b \geq 0.05$, i.e. the baryonic fraction of a halo
just scales linearly with $\Omega_b$.  Interestingly, 
the average baryonic fraction of halos in these models is greater
than the universal fraction.  Thus, somehow baryons are managing to
collapse more readily than the dark matter,
indicating that collapse beyond the boundaries of dark
matter halos is not purely gravitationally driven.  We examine this
topic in Gardner \ea\ (2001b).  The $\Omega_b=0.02$
model, however, does not have an overabundance of baryons,
showing that the process is nonlinear below some mass or density threshold.

Larger numbers of baryons within dark matter halos do uniformly lead to
enhanced cooling and collapse, as evidenced by the lower panels in
Figure~\ref{fig:baryonfrac}.  Hence, more baryons in the Universe
cause a greater fraction of them to become galactic CCGS.  We will examine the
effect of this on galaxy properties in the next section.

\begin{figure}
\vglue-0.65in
\plotone{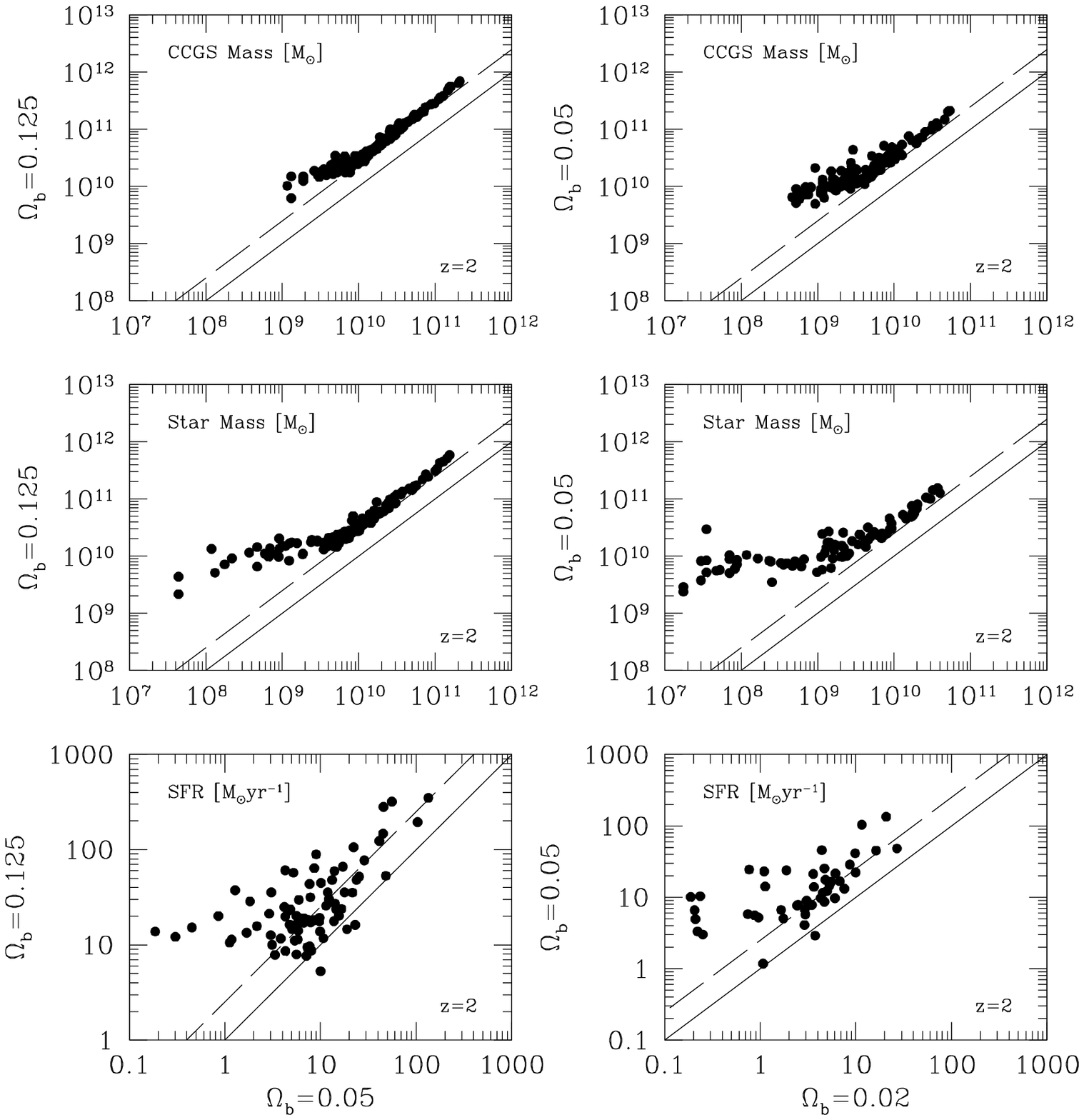} \\
\vglue-0.2in
\caption{
Similar to Figure~\protect\ref{fig:baryonfrac}, but compares total
FOF-identified halo mass in the form of CCGS (top) and stars (middle),
as well as the total SFR in each halo.
All plots are for redshift $z=2$.  The solid line shows 1-to-1
correspondence while the long-dashed line is offset by the ratio of
$\Omega_b$.}
\label{fig:matchplot}
\end{figure}

Next, let us turn our attention to the CCGS.  The upper panel of
Figure~\ref{fig:matchplot} is consistent with
Figure~\ref{fig:baryonfrac} in that both sets of points lie above the
dashed line, indicating a greater than linear dependence of CCGS mass
on $\Omega_b$.  The scaling goes as $\Omega_b^{1.23 \pm 0.02}$ in the
redshift range $2 \leq z \leq 4$ (only the $z=2$ data are shown in
Figure~\ref{fig:matchplot}).  More CCGS also produces more stars (as
opposed to just cold collapsed gas), as evidenced by the middle panels.  
Stellar mass
scales similarly to CCGS as $\Omega_b^{1.33 \pm 0.3}$.  Both scalings
are listed in Table~\ref{tab:halofits}.

Most likely, there is a critical gas mass required in the central regions
of our simulated galaxies for star formation to occur efficiently.  The
horizontal features at low halo stellar masses indicate that star
formation is severely damped at these masses for lower
$\Omega_b$ when compared to their higher $\Omega_b$ counterparts.  All
of the halos plotted in Figure~\ref{fig:matchplot} are above our
resolution limit of 64 dark matter particles.  Even so, GKHW find
similar behavior when comparing higher resolution simulations to
lower resolution ones of the same cosmology.  Consequently, this is
most likely a resolution effect.  Therefore, the scalings in
$\Omega_b$ reported above are calculated using only those points in
the higher mass regions that are well approximated by a power law (for
the top and middle panels for masses greater than $10^{10} M_\odot$;
for the bottom panels for star formation rates greater than 
0.85 M$_\odot$ yr$^{-1}$).
One may note that the scalings given here are different from those in
Section~\ref{ssec:global}.  Here we include only baryons in
dark matter halos, whereas before we examined all baryons.
We believe that the steeper scaling of the global
quantities could also be due to a resolution artifact. 
Halos at the limit of our resolution cross a critical threshold when
the gas particles are more massive, \ie higher $\Omega_b$, but do not
cross that threshold when the gas particles are less massive,
\ie lower $\Omega_b$.  Remember that the number of gas particles is
the same for all the simulations; we change $\Omega_b$ by altering the
gas particle mass.

Although the star formation rate is not shown for redshifts $z=3,4$,
the $\Omega_b$ scaling does depend strongly on redshift, going as
$\Omega_b^{1.2}$, $\Omega_b^{1.1}$, and $\Omega_b^{1.5}$ for redshifts 
2, 3, and 4, respectively. Owing to the large scatter,
it is difficult to achieve the precision of the stellar and
CCGS mass measurements.  However, it is clear that the star
formation rate scales beyond simple linearity.
Given that that the relation is close to linear for $z \leq 3$, the
increased stellar mass in the middle panels most likely comes from a
higher star formation rate at earlier redshifts.

The careful reader will note that there are points below $10^9
M_\odot$ for the $\Omega_b=0.05$ run in the left panel in the middle
row but not in the right panel.  This is because the halos in the
$\Omega_b=0.05$ run with such a low stellar mass have no
SKID-identifiable groups at all in the $\Omega_b=0.02$ run and hence
are not included in the $\Omega_b=0.05$ vs.\ $\Omega_b=0.02$ plot.
Since gas particle mass scales with $\Omega_b$, we {\em would} expect
to detect SKID groups in both simulations if stellar mass scaled
simply with $\Omega_b$.  The fact that there are no SKID groups in
the low $\Omega_b$ halos that have SKID groups at higher baryon abundance
again indicates a steep $\Omega_b$ dependence.

\subsection{Galactic Gas and Star Formation Rate}

\begin{figure}
\vglue-0.65in
\plotone{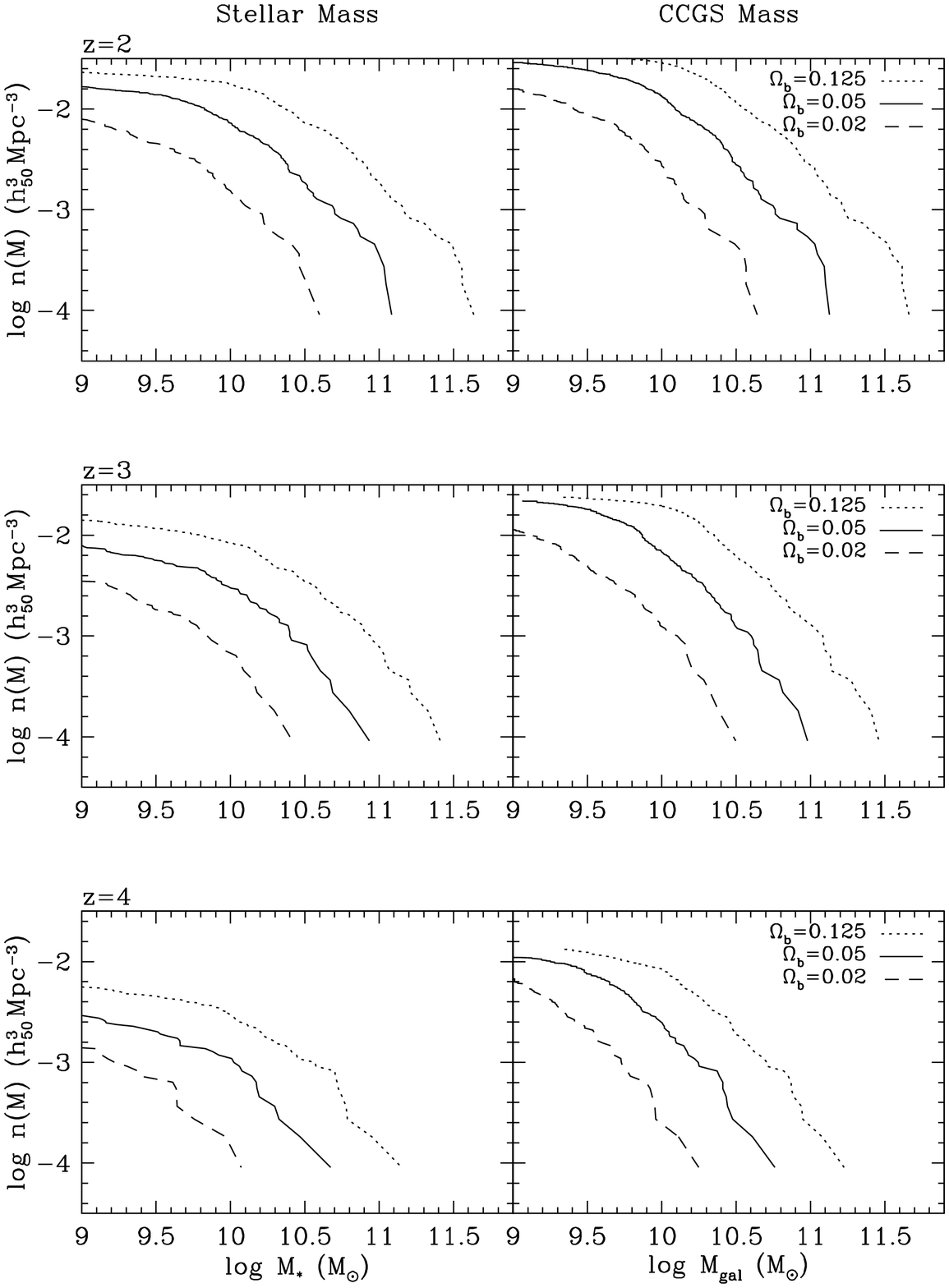} \\
\vglue-0.2in
\caption{ Effect of $\Omega_b$ on galaxy stellar and total mass.  The
lines indicate the cumulative mass function of stars (left panels) or
gas + stars (right panels) in SKID-identified groups, \ie galaxies, in each 
of the three
simulations.  }
\label{fig:stargassmass}
\end{figure}

\begin{figure}
\vglue-0.65in
\plotone{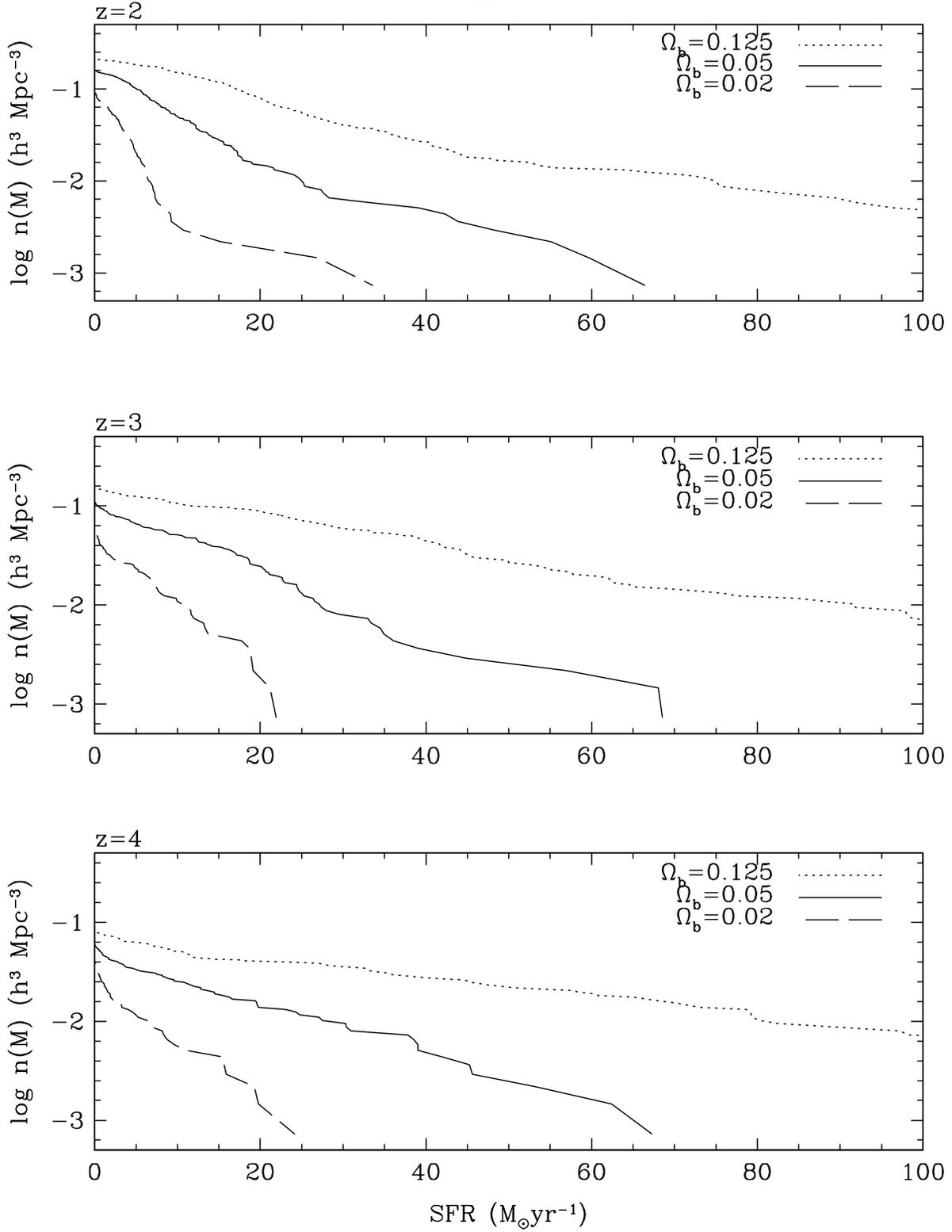} \\
\vglue-0.2in
\caption{
Effect of $\Omega_b$ on galaxy star formation rate.  The lines indicate the
cumulative number density of SKID-identified groups with a star
formation rate greater than the $x$-axis value.}
\label{fig:starform}
\end{figure}

\begin{figure}
\vglue-0.65in
\plotone{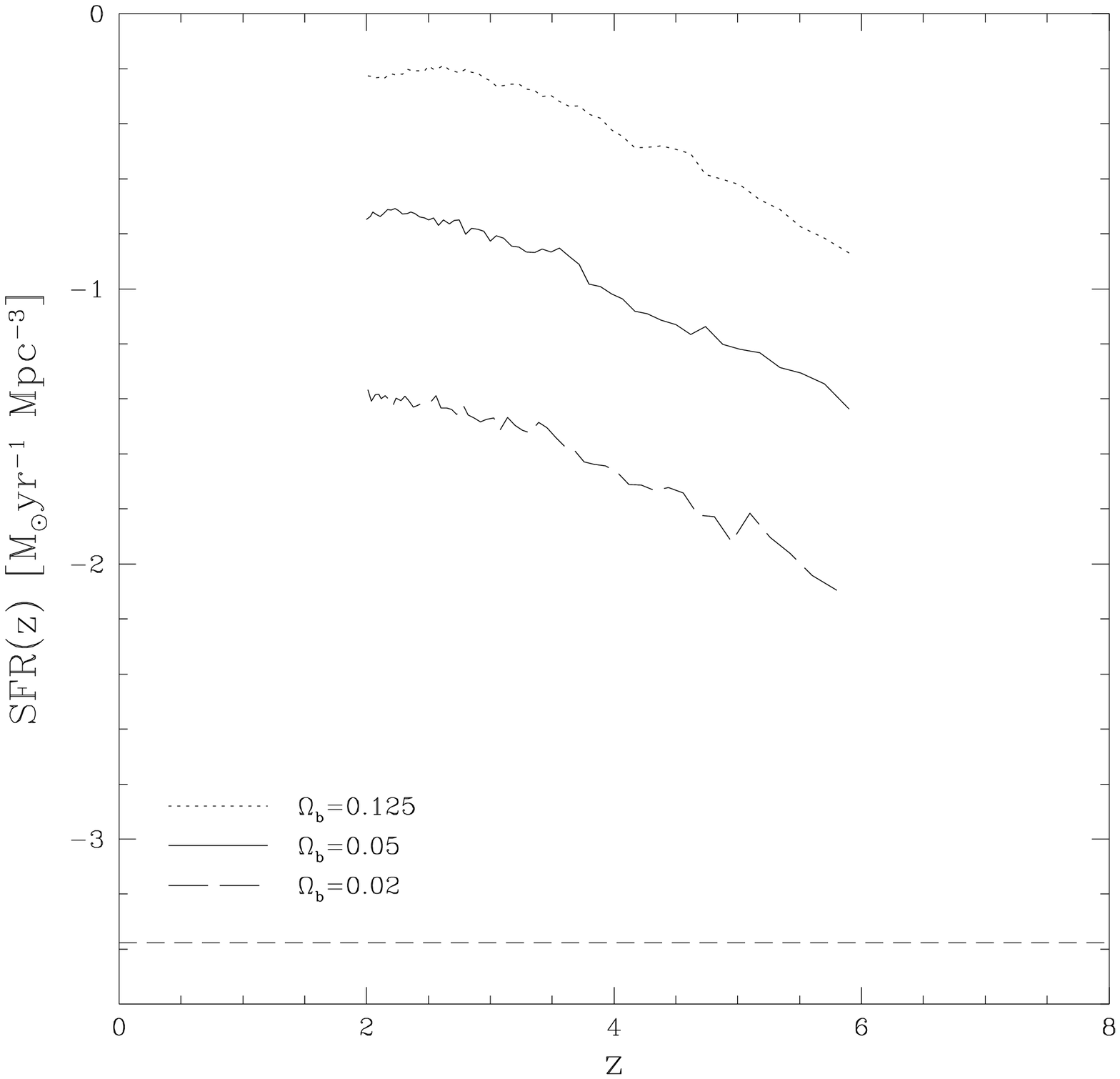} \\
\vglue-0.2in
\caption{ Total star formation in resolved halos for each $\Omega_b$
expressed as a comoving density and plotted in log space.  The
straight long-dashed line is the present mass density of metals
divided by present age of the Universe, taken from Madau et al
(1996).}
\label{fig:madauplot}
\end{figure}

In this section, we concentrate on the effect of universal baryon
fraction on galaxies.  In contrast to the previous sections, here we
consider SKID-identified groups of gas and stars --- \ie galaxies in
our simulations --- irrespective of their parent FOF-identified parent
halos.  We keep plotted values in units unnormalized to the universal
fraction to more easily relate them to observables.
Figure~\ref{fig:stargassmass} plots the cumulative comoving number
density of galaxies $n(M)$ with mass $> M_*$ in stars or $> M_{\rm gal}$ in
CCGS, while Figure~\ref{fig:starform} plots the cumulative number of
galaxies with at least a given star formation rate (SFR).  Given that
the halo mass function is essentially identical in all three
simulations, the plots indicate a direct correspondence between
$\Omega_b$ and typical galaxy gas, stellar mass and star formation
rate.  Figure~\ref{fig:madauplot} gives the total star formation rate
in resolved galaxies as a function of redshift for all three
simulations.  Note that this does not include star formation in
galaxies with $v_c \simlt 100$ \kmpersec, which make a significant contribution
to the total star formation rate at high redshift.  
The global SFR is a strong function of the universal baryon fraction and is 
well approximated by a power law.  The detailed scaling is given in
Table~\ref{tab:fits} and goes roughly as $\Omega_b^{1.5}$ with
modest redshift dependencies.

\begin{table}
\begin{tabular}{llll}
\tableline\tableline
\multicolumn{1}{c}{ }&\multicolumn{3}{c}{$\beta$} \\
 & $z=2$&$z=3$&$z=4$ \\ \cline{2-4}
Star Formation Rate &1.2	&1.1	&1.5 \\
CCGS		&\multicolumn{3}{c}{$1.23 \pm 0.02$} \\
Stars		&\multicolumn{3}{c}{$1.33 \pm 0.03$} \\
\tableline\tableline
\end{tabular}
\caption{ Table of baryonic dependence, $\Omega_b^\beta$, of CCGS, stars, and
star formation rate inside dark matter halos individually identified in
each simulation, as given in Figure~\protect\ref{fig:matchplot}.}
\label{tab:halofits}
\end{table} 

\section{Summary}

We present an analysis of the effects of $\Omega_b$ on a broad range of
measurable properties in our simulations.  Our results our summarized in
Tables 1--3, where we present the power law scaling of the quantities with
$\Omega_b$.  The amount of baryonic material scales linearly with
$\Omega_b$.  The cooling rate scales as $\Omega_b^2$, so one might have
expected the scalings of quantities related to the condensed galactic
material to scale with a power law index between one and two, and that is
just what we find.  

Dividing the total gas in
the each simulation into three phases, based on density and temperature,
we find that the primordial diffuse phase remains largely
unaffected by changes in $\Omega_b$.  Shocked and condensed gas,
however, exchange abundances with condensed gas being more prevalent
in higher $\Omega_b$ universes while less mass is contained in the
shocked phase.  This is consistent with Murali et al. (2001) who find that
galaxies gain mass mostly through the smooth accretion of cooling gas and
not through merging.  A general theme of our results is that higher
$\Omega_b$ promotes increased gas cooling.  The mass in condensed gas
and stars is well described by a power law in $\Omega_b$ of
greater than linear scaling, as detailed
in Table~\ref{tab:fits}.  Dark matter halos in
higher $\Omega_b$ cosmologies contain more than their fair share of
both cold collapsed gas and stars, meaning galaxies also tend to be
more massive at $2 \leq z \leq 4$ in these universes.  A larger
fraction of galaxies in high $\Omega_b$ models have high rates of star
formation, and the universal star formation rate is also greater.
Cold collapsed gas and stars as well as stellar mass alone scale
superlinearly as given in Table~\ref{tab:halofits}.

Because it probes gas in the diffuse phase, the LAF is insensitive
to the value of $\Omega_b$ provided that the UV background intensity
is adjusted to reproduce the observed mean opacity of the forest.
The required UV background intensity scales approximately as
$\Omega_b^{1.7}$, and diffuse gas in a higher $\Omega_b$ model
is slightly hotter because of the higher photoionization rate.
The gas overdensity field is essentially independent of $\Omega_b$
in diffuse regions, and higher neutral fractions compensate for
lower gas densities when $\Omega_b$ is low, so synthetic spectra
are virtually identical for different values of $\Omega_b$.
These results support the standard practice of scaling LAF spectra
from hydrodynamic simulations to match the observed mean opacity.

As one would intuitively expect, models with higher $\Omega_b$ have
increased DLA and LL absorption.  The $\Omega_b$ dependence is much
stronger for DLA absorption than for LL absorption, and it is somewhat
stronger at $z=2-3$ than at $z=4$.  Our simulations only resolve
baryon concentrations in halos with $v_c \geq 100\kms$, and they
therefore underestimate the total incidence of DLA and LL absorption.
However, the trend of increased absorption with increasing $\Omega_b$
holds on a halo-by-halo basis within the mass range that we do 
resolve, so we expect that it would continue to apply to the 
absorption in all halos.

Taken together, our results imply a relative simple picture of the
influence of $\Omega_b$ on high-redshift structure.  The amount of
gas in the diffuse phase is essentially unaffected by changes in
$\Omega_b$, and the LAF absorption produced by this gas is unchanged
if the UV background is adjusted to keep the mean opacity fixed.
The increased cooling in models with high $\Omega_b$ shifts gas
from the shocked phase to the condensed phase, increasing the 
stellar and gas masses of galaxies, the rates of star formation, 
and the amount of high column density absorption.  
While the $\Omega_b$-scalings of these quantities that are 
listed in Table~\ref{tab:fits} are not guaranteed to hold in 
other cosmologies or other ranges of redshift and galaxy mass,
they provide a best guess for how to scale the predictions of
hydrodynamic simulations to alternative values of $\Omega_b$.
If the current consensus on the value of $\Omega_b$ survives
improvements in the observations, then the remaining uncertainties
in $\Omega_b$ will contribute relatively little uncertainty
to predictions of high-redshift structure.

\acknowledgements

We thank Mark Fardal for useful discussions and input during the data
analysis.  This work was supported by NASA Astrophysical Theory Grants
NAG5-3922, NAG5-3820, and NAG5-3111, by NASA Long-Term Space
Astrophysics Grant NAG5-3525, and by the NSF under grants ASC93-18185,
ACI96-19019, and AST-9802568.  Gardner was supported by NASA Grant
NGT5-50078 and NSF Award DGE-0074228 for the duration of this work.
The simulations were performed at the San Diego Supercomputer Center.

\vfill\eject


\begin{thebibliography}{}

\bibitem[Barnes \& Hut(1986)]{bh86}
Barnes, J.E. \& Hut, P. 1986, Nature, 324, 446

\bibitem[Burles \& Tytler(1998ab)]{burles98ab}
Burles, S., \& Tytler, D. 1998a, \apj, 499, 699

\bibitem[Burles \& Tytler(1998)]{burles98b}
Burles, S., \& Tytler, D. 1998b, \apj, 507, 732

\bibitem[Croft et al.(1997)]{croft97}
Croft, R.A.C., Weinberg, D.H., Katz, N., Hernquist, L., 1997, \apj, 488, 532

\bibitem[D'Odorico et al.(2001)]{dodorico01} D'Odorico, S.,
Dessauges-Zavadsky, M., \& Molaro, P. 2001, \aap, in press, astro-ph/0102162

\bibitem[Dav\'e et al.(1999)]{Dave99}
Dav\'e, R., Hernquist, L., Katz, N., \& Weinberg, D.H. 1999, \apj,
511, 521.

\bibitem[Gardner et al.(2001a)]{gkhw99}
Gardner, J. P., Katz, N., Hernquist, L., \& Weinberg, D. H. 2000a,
\apj, in press, astro-ph/9911343 (GKHW)

\bibitem[Gardner et al.(2001b)]{gkwh01}
Gardner, J. P., Katz, N., Weinberg, D. H., \& Hernquist, L. 2000b, in preparation

\bibitem[Gardner et al.(1997)]{gkhw97}
Gardner, J. P., Katz, N., Hernquist, L., \& Weinberg, D. H. 1997,
\apj, 484, 31

\bibitem[Gingold \& Monaghan(1977)]{gingold77}
Gingold, R.A. \& Monaghan, J.J. 1977, \mnras, 181, 375

\bibitem[Haardt \& Madau(1996)]{HM96} 
Haardt F., \& Madau P. 1996, \apj, 461, 20 

\bibitem[Hernquist(1987)]{h87}
Hernquist, L. 1987, \apjs, 64, 715

\bibitem[Hernquist \& Katz(1989)]{hk89}
Hernquist, L. \& Katz, N. 1989, \apjs, 70, 419

\bibitem[Hernquist et al.(1996)]{HK96}
Hernquist, L., Katz, N., Weinberg, D.H. \& Miralda-Escud\'e, J.
           1996, \apj, 457, L51

\bibitem[Jaffe et al.(2001)]{maxima} Jaffee, A.H., et al.\ 2001, \prl, 86, 3475 

\bibitem[Katz, Hernquist \& Weinberg(1999)]{khw99} 
Katz, N., Hernquist, L., \& Weinberg D.H. 1999, ApJ in press, astro-ph/9806257

\bibitem[Katz \& Quinn(1995)]{TIPSY} 
Katz, N., \& Quinn, T. 1995, TIPSY manual

\bibitem[Katz, Weinberg, \& Hernquist(1996)]{kwh96} 
Katz, N., Weinberg D.H., \& Hernquist, L. 1996, \apjs, 105, 19 (KWH)

\bibitem[Kurki-Suonio \& Sihvola(2001)]{KS00} Kurki-Suonio, 
H.\ \& Sihvola, E.\ 2001, \prd, 63, astro-ph/0011544

\bibitem[Lange et al.(2000)]{boomerang} 
Lange, A. E., and 31 colleagues 2000, astro-ph/0005004

\bibitem[Lucy(1977)]{lucy77}
Lucy, L. 1977, \aj, 82, 1013

\bibitem[Miralda-Escud\'e et al.(1996)]{miralda96}
Miralda-Escud\'e J., Cen R., Ostriker, J.P., \& Rauch, M. 1996, \apj, 471, 582

\bibitem[Murali et al. (2001)]{murali01}
Murali, C., Katz, N., Hernquist, L., Weinberg, D. H. \& Dave\'{e}, R. 2001,
astro-ph/0106282

\bibitem[Netterfield et al.(2001)]{netterfield01} Netterfield, C.\ 
B.\ et al.\ 2001, \apj, submitted, astro-ph/0004460

\bibitem[O'Meara et al.(2001)]{omeara01} O'Meara, J.M., Tytler, D.,
Kirkman, D., Suzuki, N., Prochaska, J.X., Lubin, D., \& Wolfe,
A.M. 2001, \apj, in press, astro-ph/0011179

\bibitem[Padmanabhan \& Sethi(2000)]{PS00} Padmanabhan, T., \& Sethi,
S.K. 2000, \apjl, submitted, astro-ph/0010309

\bibitem[Pettini \& Bowen(2001)]{PB01} Pettini, M., \& Bowen,
D.V. 2001, \apj, submitted, astro-ph/0104474

\bibitem[Press, Rybicki, \& Schneider(1993)]{PRS}
Press, W. H., Rybicki, G. B., \& Schneider, D. P. 1993, \apj, 414, 64

\bibitem[Pryke et al.(2001)]{pryke01} Pryke, C., Halverson, N.\ 
W., Leitch, E.\ M., Kovac, J., Carlstrom, J.\ E., Holzapfel, W.\ L., \& 
Dragovan, M.\ 2001, \apj, submitted, astro-ph/0004490 

\bibitem[Rugers \& Hogan(1996)]{rh96} Rugers, M. \& Hogan, C. 
J. 1996, \apjl, 459, L1 

\bibitem[Songaila, Cowie, Hogan \& Rugers(1994)]{songaila94} 
Songaila, A., Cowie, L. L., Hogan, C. J. \& Rugers, M. 1994, \nat, 368, 599 

\bibitem[Stadel et al.(2000)]{skidpaper}
Stadel, J., Katz, N., Weinberg, D.H., \& Hernquist,
           L. 2000, in preparation

\bibitem[Storrie-Lombardi et al.(1996)]{SL96}
Storrie-Lombardi, L.J., Irwin, M.J., \& McMahon, R.G. 1996,
           \mnras, 282, 1330

\bibitem[Storrie-Lombardi et al.(1994)]{SL94}
Storrie-Lombardi, L.J., McMahon, R.G., Irwin, M.J., \&
           Hazard, C. 1994, \apj, 427, L13

\bibitem[Storrie-Lombardi \& Wolfe(2000)]{slw00}
Storrie-Lombardi, L.\ J.\ \& Wolfe, A.\ M.\ 2000, \apj, 543, 552

\bibitem[Tytler et al.(2000)]{tytler00} Tytler, D., O'Meara, J.M.,
Suzuki, N., \& Lubin, D.\ 2000, \physscr, in press, astro-ph/0001318

\bibitem[Webb et al.(1997)]{webb97}
Webb, J.\ K., Carswell,
R.\ F., Lanzetta, K.\ M., Ferlet, R., Lemoine, M., Vidal-Madjar, A., \& 
Bowen, D.\ V.\ 1997, \nat, 388, 250 

\end{thebibliography}
\end{document}